# Mechanisms of nearshore retention and offshore export

# of mussel larvae over the Agulhas Bank

*Authors:* Nicolas Weidberg[1], Francesca Porri[1, 2], Charles von der Meden[1,3], Jennifer M. Jackson[4], Wayne Goschen[3], Christopher McQuaid[1]

1 Department of Zoology and Entomology, Rhodes University, Grahamstown 6140, South Africa

2 South African Institute for Aquatic Biodiversity, Somerset Street, Grahamstown 6139, South Africa

3 South African Environmental Observation Network, Egagasini Node, Martin Hammerschlag Way, Roggebaai 8012, South Africa

4 ASL Environmental Sciences Inc., Rajpur Place, Victoria V8M 1Z5, British Columbia, Canada.

Correspondence: Nicolas Weidberg, N.Weidberg@ru.ac.za

**Abstract:** Ecological connectivity is critical for population dynamics but in many benthic species it is complicated by a planktonic larval phase, whose dispersal remains poorly understood. Using a plankton pump, we examine the distribution of intertidal mussel larvae along three axes: alongshore, cross-shelf and by depth during a large scale (600km) cruise over the Agulhas Bank off southern Africa in August/September 2010. As a general pattern, higher veliger abundances were found close to the coast. Our analyses of the nearshore flow, estimated from ADCP data and the vertical distribution of larvae, show that onshore larval retention may be mediated by active vertical swimming through the water column guided by light and wind-induced turbulence. A massive offshore export of larvae off St Francis Bay was, however, observed during an Agulhas Current meander which influenced inner shelf waters. We hypothesize that, by increasing and homogenizing flow, the Agulhas Current may erase the effects of larval vertical positioning on onshore retention and transport larvae offshore. Our study highlights the need to integrate the effects of complex, region-specific physical dynamics with the swimming behaviour of larvae in order to explain their spatial distribution, population connectivity and the consequences for population dynamics.

# INTRODUCTION

Genetic and ecological connectivity of marine systems depend largely on the dispersal of planktonic propagules, whose dispersal patterns are difficult to identify given the range of hydrographical mechanisms and behaviours that influence larvae across varying spatial scales (Palmer and Strathmann, 1981; Levin, 1992). The problem is even more complicated when applied to intertidal, relatively isolated, populations of sedentary organisms that release planktonic larvae into coastal waters characterized by highly stochastic flow regimes (Botsford et al.; 2001, Cowen et al., 2006; Pineda et al., 2007; Pineda et al., 2010).

Traditionally, planktonic larvae were thought to be passive in energetic environments, which results in extensive dispersal, on the order of hundreds of kilometers (Thorson, 1950; Thorson, 1961; Scheltema, 1971; Roughgarden et al., 1985; Roberts and Mullon, 2010). However, clear evidence of onshore retention and restricted dispersal has been observed in several coastal systems for intertidal invertebrate larvae (McQuaid and Phillips, 2000; Poulin et al., 2002; Shanks and Brink, 2005; Morgan et al., 2009; Shanks and Shearman, 2009; Morgan et al., 2011). In essence, two theories, not mutually exclusive, have been proposed to explain onshore larval retention patterns in the dynamic coastal environment. Passive retention processes are possible because a horizontal boundary layer is often seen along the coast (Csanady, 1972; Wolanski, 1994; Largier, 2003). This boundary layer, typically restricted to the first few kilometers off the coast, would increase the retention of particles (Wolanski and Spagnol, 2000; Largier, 2003). Larvae cannot overcome passive retention processes by swimming. Alternatively, active retention may result from vertical swimming behaviour. By migrating through the water column, even planktonic organisms with no swimming-related appendages may slow down offshore transport. This has been demonstrated for surf zone diatoms, which alter their buoyancy by attaching to air bubbles during the day and floating to the surface (Talbot and Bate, 1987; Talbot et al., 1990).

For active larvae, three physical cues may trigger active swimming: light, wind-driven vertical flows and tide. Diurnal changes in light cause some organisms to migrate vertically, which may allow them to avoid both predation and offshore export (Marta-Almeida et al., 2006). Wind-driven Ekman circulation causes upwelling and downwelling, and swimming by the larvae of benthic invertebrates in response to this changing flow helps them stay near the coastline (Poulin et al., 2002; Genin et al., 2005; Shanks and Brink, 2005; Shanks and Shearmann, 2009). Larvae may also to react to turbulence, thus escaping from offshore advection by sinking below the Ekman layer (Franks, 2001; Pringle, 2007). Finally, tides influence the dynamic structure of the water column in some systems including wide, shallow shelves such as the Agulhas Bank in South Africa (Schuman and Perrins, 1982). Under such physical forcing, swimming in response to vertical currents may allow larvae to be retained onshore (Epifanio and Garvine, 2001; Queiroga et al., 2006).

Although understanding coastal physical dynamics and active larval swimming behaviour may help explain larval distribution, specific characteristics of an organism or a system can have significant effects on larval connectivity. For example, mussel larvae are known to be slow swimmers (a few mm s$^{-1}$, Chia et al., 1984) so the initial assumption of passivity in the water column cannot be ruled out. Previous studies have shown that mussel veligers are distributed homogenously through the water column (Porri et al., 2014) and drift passively with the winds (McQuaid & Phillips, 2000).The larval pool found in those studies, however, was made up of relatively small patches in nearshore waters (within a few hundreds of meters of the coastline), which does not match the much larger scales of passive diffusion in the coastal ocean (Pineda et al., 2007). Moreover, the horizontal scales at which these larvae are dispersed and recruited are only of the order of tens of kilometers (McQuaid and Phillips, 2000; Lawrie and McQuaid 2001; Porri et al., 2006, 2008), making active mechanisms of retention a realistic possibility.

Here we study the Agulhas Bank, an extension of the continental shelf at the southern end of Africa (Figure 1). The Agulhas Bank is influenced by the Agulhas Current, which runs southwestwards parallel to the coastline at speeds greater than 1m s$^{-1}$ (see Lutjeharms, 2006 for review). The Agulhas current normally follows the continental shelf break along the 200 m isobath, however, periodic solitary meanders (also called large episodic meanders) can cause the current to be displaced (Grundlingth, 1979).. The leading edge of the meander causes the current to be displaced onto the continental shelf in a feature that is called an Agulhas Current Meander (Jackson et al., 2012). The lagging end of the solitary meander causes the current to be displaced offshore by about 100 km.  Agulhas Current meanders cause strong currents over the continental shelf and also advect phytoplankton (Jackson et al., 2012) and mussel larvae (Porri et al., 2014) offshore from the shelf within the current. A recent analysis of satellite data suggests that solitary meanders impact the circulation of the eastern Agulhas Bank for about 110 days per year (Krug et al., 2014). The effects of a solitary meander on plankton distribution are likely to alter the way larvae interact with, and respond to wind, tidal and light-mediated processes.  During a winter 2010 cruise, physical and larval data were collected both before and as a solitary meander passed, offering a unique opportunity to test the responses of mussel veligers to unusual forcing (the solitary meander) and normal forcing (light, wind, and tides).

We used a large scale grid of stations, one of the most complete and extensive of its kind, to analyse spatial patterns of mussel larvae over the Agulhas Bank. As both physical structures (Jackson et al., 2012) and large scale larval abundances (Porri et al., 2014) have been described we focus on the finer scale descriptors of larval distribution: mean larval depth and median larval distance offshore. These descriptors represent distances through the water column and across the shelf, indicating where the core of the larval population is located, regardless of its magnitude, thus avoiding the issue of stochastic alongshore variability in the patterns of abundance that cannot be explained without temporal replication (Lagos et al., 2008; Ayata et al., 2012). Instead, our approach allows detailed inspection of larval abundances according to depth and cross-shelf position in relation to general hydrodynamics driven by wind and tide as well as light and the forcing of the Agulhas Current. In addition, we incorporate simple advection-diffusion Gaussian models based on moored ADCP data for comparison with real larval distributions.

MATERIALS AND METHODS

The Agulhas Bank cruise took place from 31 August to 13 September 2010 from Cape Padrone to Cape Agulhas along approximately 600km of coastline. A total of 110 stations were sampled along 12 transects, with 5-16 stations per transect, covering the entire width of the Agulhas Bank. Within each transect, stations were separated by an average of 10.11±1.75 (SD)km and the average alongshore distance between transects was 44.7±11.74km (Fig. 1).

*Environmental variables:* At each station, temperature and salinity profiles were obtained using a SBE 911 Plus CTD and photosynthetically active radiation (PAR) profiles were obtained using a Satlantic PAR sensor. A ship-based RD instruments OS75-I-2-UG 75 kHz ADCP current meter measured currents with a vertical resolution of 8m. ADCP measurements started at 35 m so there was no current data from the near-surface waters.

Wind speed and direction were obtained from 5 meteorological stations along the coast (Fig. 1) from 15 August to 15 September 2010. These stations cover the whole Agulhas Bank, extending from Cape Infanta to Cape Padrone (Fig. 1). To obtain representative data, hourly wind measurements for each meteorological station were assigned to the closest transects sampled: Port Elizabeth for transects 1, 2 and 3; Cape St Francis for transects 4, 5 and 6; Plettenberg Bay for transects 7 and 8; Mossel Bay for transects 9, 10 and 11; and Struisbaai for transect 12. Anemometers were located on average 69±66m above the ground. Because this wide range in height may introduce variability wind data were corrected with the wind profile power law, following the equation:

$v_{9m} = v_h * (9/h)^{0.11}$  \hfill 1)

where $v_{9m}$ is the wind velocity at a height of 9m (altitude at the Cape St Francis´ weather station) and $v_h$ is the wind velocity measured at each station at height $h$ (see Hsu et al. 1994 for parameter estimations).

Time series of wind data from Port Elizabeth (15 August- 15 September 2010) were inspected to determine the relevant time scales at which upwelling and wind driven turbulence (see below) fluctuated significantly. Wind velocity data were analyzed using a Fourier transform to obtain the periodograms. Unlike other statistical procedures, periodograms provide the variability explained by different periods, thereby revealing the time scales at which a given variable varies significantly. Tidal amplitude (TA) was taken from tide tables for Port Elizabeth (data used for transects 1 to 6) and Mossel Bay (transects 7-12).

Daily sea surface temperature (SST) satellite imagery (http://www.remss.com/sst) was used to characterise mesoscale hydrographic processes occurring during the survey, using the maximum inshore point in the SST gradient to approximate the position of the Agulhas Current.

*Larval abundances:* Plankton were collected at 78 stations using a 2.2 KC Denmark 23.580 plankton pump with a net of 60µm mesh. At each station, three different depths were sampled: surface (1-5m depth), thermocline (indicated by the CTD profiles) and below the thermocline to the maximum length of the plankton pump cable (approx. 200 m). The total volume filtered ranged from 1500 to 2000L during 10 min with a flow rate of 30 cm/s. Samples were preserved in 70% ethanol. Mytilid veligers smaller than 200µm were identified and counted using a stereoscopic dissecting microscope. At these early stages, mytilids can be distinguished from other bivalves but species identification may not be possible using larval morphology (Porri et al. 2014). When larval densities were high, aliquots obtained with a Motodo plankton splitter (Motodo, 1959) were analyzed.

Mean depth distributions (MDD) were calculated for all stations where larvae were found in significant amounts (>10 individuals.m$^{-3}$) at at least one of the three depths. Mean depths were obtained as a weighted average following Tapia et al. 2010:

$$MDD=(\sum N_i * D_i) / (\sum N_i) \qquad \qquad 2)$$

where $N_i$ stands for the larval abundance at a depth $D_i$ and the sums comprise the three depths sampled per station.

By using mean larval abundances at every station (average over the three depths) for each transect, median larval distance offshore (MLDO) was obtained as the distance across the transect where median larval concentrations were located. This median was preferred over the average distance offshore (Shanks and Shearmann, 2009) as, in our case, this last parameter was strongly influenced by outliers and did not reflect the position of the core of the larval population across the shelf.

*Regression analyses:* To explain the spatial patterns of larvae through the water column and across the shelf, MDD and MLDO were submitted to statistical analyses as dependent variables. Because larval vertical positioning may be guided by light, tide and wind, four different predictors were used to characterise MDD: surface photosynthetically active radiation (SPAR), tidal amplitude (TA), upwelling index (UPW – see below) and wind induced turbulent energy ($\varepsilon$; see Table 1 for all the acronyms). As coastal upwelling mostly affects inshore stations, UPW was calculated only for the three stations nearest the shore (mean maximum distance offshore = 23.86±4.93km). Both UPW and $\varepsilon$ were calculated using hourly wind measurements from the meteorological stations. Positive values of the upwelling index ($m^3$ $km^{-1}$ $s^{-1}$) indicate positive Ekman transport and upwelling, while negative values point to downwelling (Bakun, 1976). UPW was calculated as follows:

$$UPW = \rho_a * C_D * v_{9m} * v_{9m-x} * f^{-1} * \rho_w^{-1} \qquad 3)$$

where $\rho_a$ stands for air density (1.22kg $m^{-3}$), $C_D$ is the drag coefficient approximated as 0.0014, $v_{9m}$ is mean height-corrected wind speed, $v_{9m-x}$ its alongshore vectorial component (estimated as zonal winds for our coast), $f$ is the Coriolis parameter ($9.9*10^{-5}$ at middle latitudes) and $\rho w$ is the estimated water density (1025kg $m^{-3}$).

$\varepsilon$ measures the dissipation rate of turbulent kinetic energy in a viscous environment and was obtained as (Pringle, 2007):

$$\varepsilon = (v_{9m}^*)^3 k^{-1} z^{-1} \qquad 4)$$

Where $v_{9m}^*$ is the height-corrected velocity scale of the turbulence (typically one thousandth of that of the wind), k is Von Karmen´s constant (0.41), z is depth (fixed at 1m for our calculations) and $\varepsilon$ is measured in W $kg^{-1}$. As the time scale at which winds may influence larval distributions was not known *a priori*, UPW and $\varepsilon$ were calculated from wind data averaged over 8 periods (12h and 1, 2, 3, 4, 5, 6 and 7 days prior to sampling). Shorter periodicities were not considered as on smaller time scales than the inertial period (around 17h) significant movement of the water masses cannot be attained by wind forcing (Pringle, 2007). These estimates were introduced separately in a multiple linear regression analysis as predictors together with TA and SPAR, with MDD as the dependent variable. Linear dependency between predictors was not allowed, to avoid multicollinearity. In those models presenting more than one significant predictor, the interaction term between them was calculated and added to the model to further test its significance. Models were selected and ordered using the Akaike Information Criterion which takes into account the amount of variance explained but penalizes the complexity of the model (see Burnham and Anderson 2002 for review).

As the Agulhas Currents may be the main physical agent driving offshore advection of larvae, the distance at which the bulk of the larval population was located across the shelf at each transect (MLDO) was regressed against the distance at which the inner border of the Agulhas was found. This distance was estimated using both the currents measured with the ship-based ADCP and surface sea temperature thresholds (Lutjeharms, 2006). The temperature values that approximate the Agulhas Current range from 21ºC to 17ºC from the longitude of Algoa Bay to that of Cape Agulhas in winter (Christensen, 1980). When the current was offshore of the shelf break, satellite imagery was used to find the edge of the current. The spatio-temporal scales at which the Agulhas Current may influence larval distributions are unknown. Thus, MLDO at a given transect was regressed with the distance from the coast to the current at that same transect but also at transects located upstream (eastwards). This approach accounts for both spatial and temporal lags because consecutive transects were separated by an average distance of 44.7±11.74km and a time period of about 1 day. As multiple tests are performed with the same dependent variable, we applied the false discovery rate correction (FDR, Benjamini and Hochberg, 1995) to the raw p-values obtained, thus reducing the probability of incurring a Type I error. The FDR defines a new, more conservative alpha value based on the number and significance of the models tested (García, 2004).

*Theoretical vs real cross shelf larval distributions.* - To determine whether nearshore distributions of larvae were driven by passive retention processes or active larval behaviour,, real and theoretical spatial patterns were compared. For this exercise, data from the moored ADCP (Teledyne RDI Express 43.03 600 kHz) current meter at Bird Island (Fig. 1) were used to build a theoretical mean dispersal function. Current measurements were recorded every 20 minutes with a spatial resolution of 0.5m from 6 to 28.8m deep from 1 August to 30 September 2010. As the ADCP beam angle was 20º, measurements from the first 6 meters of the water column had to be discarded (Gordon, 1996). This data set may be representative of coastal waters in the region, as the instrument was placed close to the shore at a depth of 30m and has already been used to analyse nearshore hydrography (Goschen et al., 2012). To calculate the theoretical larval distributions, an advective-diffusive model (Largier, 2003; White et al., 2010) was used expressed by the Gaussian equation:

$N=(A/(L_D (2\pi)^{1/2}))*\exp(-(D-L_A)^2/(2L_D^2))$                5)

where N stands for larval abundance (ind.m$^{-3}$), A is the amplitude of the normal curve or maximum abundance (ind.m$^{-3}$) and D represents distance from every station to the coast.. $L_A$ is the advective length, which can be inferred by multiplying the time a given larva has been in the water (T) by the mean current vector during that period (v, Largier, 2003). $L_D$ is the diffusive length, whose value is calculated using the current variability (the standard deviation σ) and the decorrelation time scale (τ), typically set at 0.5 days at the coast (Largier, 2003; Siegel et al., 2003; White et al., 2010):

$$L_D = (T * \tau * \sigma^2)^{0.5} \qquad \qquad 6)$$

As observed larvae may represent a mix of ages, $L_A$ and $L_D$ were calculated every 20min from 40 hours to 8 days before the closest transect was sampled (transect 2, 2/09/2010 at 10:00AM) setting larval departure from the shore. These time intervals were considered to comprise the planktonic larval duration of the early veligers whose abundances were analysed: from the occurrence of the first D-stage larva (40h after hatching) to the presence of the first rounded-umbo veliger after 8 days (de Schweinitz and Lutz, 1976). Based on previous research done in the region (Phillips and McQuaid, 2000; Porri et al., 2014), larvae were assumed to be evenly distributed through the water column because they have not shown a preferred depth. Thus, cross shelf currents for all depths were averaged for each time interval. Advective and diffusive lengths were calculated in this way for all periods (T= 40h; T=40h and 20 min; T= 40h and 40 min…T= 8 days, n=456). To account for larval mortality, which may affect the age structure of the larval population, a survival rate of 0.8796 day$^{-1}$ (0.9983/20min) was applied to match the observed mortality rate for *Mytilus edulis* (Jorgensen, 1981) veligers. Thus, advective and diffusive lengths were multiplied by their respective weight ($L_{A0}=L_{A\_40h}*1$, $L_{A1}=L_{A\_40h,20min}*0.9983$, $L_{A2}=L_{A\_40h,40min}*(0.9983)^2$… $L_{A456}=L_{A\_8days}*(0.9983)^{456}$ ). These mortality-corrected estimates were then averaged and the resulting means used in equation 5) to build the final theoretical distribution following Rivera et al. (2013).

It is possible that alongshore flow may introduce uncertainty, since larvae collected at transect 2 may be advected from other locations along the coast and not experience the currents measured at Bird Island. To account for this uncertainty, measured vertically-averaged alongshore currents were submitted to the same analysis applied to the cross shelf flow and an alongshore theoretical distribution was also obtained for $L_D$ and $L_A$.

To determine the empirical distribution, larval abundances at all depths for transect 2 were regressed against distance from the coast (D) using the same Gaussian fit. Hence, estimates for $L_D$ and $L_A$ which describe the real distribution of larvae were obtained. To compare real and theoretical distributions, the sum of all the larval densities predicted by the ADCP-based model from 0 to 100km offshore was restored to the observed number of larvae found across transect 2.

To account for the temporal variability of the currents, alongshore and cross-shelf components of the flow measured at Bird Island from 1 August to 30 September 2010 for the surface (6.3m), middle (12.8m) and bottom (28.8m) layers were submitted to Fourier transforms to build the corresponding periodograms. Dominant periods were compared with those time scales at which wind-induced turbulence, tidal constituents, and upwelling processes were most energetic, according to their respective Fourier transforms, and when physical processes were most correlated with larval vertical distributions.

RESULTS

*Hydrographical structure and larval distributions:* Cross shelf profiles of temperature, salinity and currents reveal different spatial patterns over the Agulhas Bank. Transects 1-4, sampled from 31 August to 3 September, were characterised by moderate to low cross shelf flow (Fig. 2). Relatively high larval abundances were found very close to the coast along transects 1-3 (MLDO at 3.7km). On transects 5 and 6, high temperatures (>21ºC) and strong currents (>1m s$^{-1}$) indicated that the Agulhas Current was on the continental shelf about 15-30 km from the coastline at transect 5 and about 33-48 km from the coastline at transect 6 on 4-5 September (Fig. 3). The larval population core was displaced offshore (MLDO at 18 and 33 km for transects 5 and 6, respectively). Transects 7-8 (6-8 Sept) were characterized by relatively fresh, cold bottom waters (35.1, 10.5ºC), northward bottom currents, and isopycnals that sloped upwards towards the coast, which suggests upwelling (Fig. 3). Larval concentrations were highest near the coast (MLDO at 3.7km, Fig. 3). At transects 9-12, (9 - 13 Sept), currents were generally slow and variable (Fig 4). Across transect 12, the direction of flow changed, and coincided with vertical heaving of the thermocline (Fig 4), suggesting the presence of internal waves(Jackson et al. 2012). Although larval abundances were always higher at nearshore stations for these transects (MLDO from 3.7 to 7.4 km offshore), veligers were also found offshore and at the surface in significant numbers 87km from the coastline along transect 12 (Fig 4).

*Regression analysis:* Three significant models from the multiple regression analysis were obtained using MDD as the dependent variable including all stations (n = 36) where larvae were present in significant numbers following Weidberg et al. 2014 (>10 ind. m$^{-3}$). The best fit was produced by a model containing surface photosynthetically active radiation (SPAR) and wind induced turbulence averaged over 2 days, which explained 37% of the total variability in the MDD data (Table 2). When the interaction term between the two best predictors was added to the model, it was not significant. The relationship between larval position and SPAR and wind induced turbulence was negative, indicating that the higher the turbulence and the irradiance, the deeper the larvae were positioned in the water column (Table 2). Regressions with SPAR and ε (2days) were also significant (Table 2, Fig. 5). Similar results were obtained for the inshore stations (n = 26) At the inshore stations, the combined model with SPAR and ε (2days) explained up to 25% of the variability in MDD without the interaction term, but separately none of these predictors were significant (Table 2). Based on AIC, the best model for these inshore stations included Ekman transport, as MDD was positively correlated with UPW-5days (Table 2, Fig. 5). Tide did not contribute significantly to any of these models (Fig. 5).

Wind data from the Port Elizabeth time-series showed very clear temporal patterns (Table 3). Periodograms peaked at 110 hours (4.6 days) for both wind speed and the upwelling index but daily winds explained most of the variability in wind induced turbulence (Table 3).

MLDO was significantly correlated with distance to the Agulhas Current at a lag of 1 transect (Fig. 6). The distances to the inshore edge of the Agulhas Current was estimated using vertical profiles of temperature and currents except for transects 1, 2, 3 and 8, where the current was offshore of the shelf and satellite imagery was required (Fig. 1, Fig. 7). For these linear regression analyses, MLDO at transect 1 was excluded as larval abundances along the whole transect were too low for all the stations and all depths (<10 ind. m$^{-3}$). The sign of the fit was negative and it remained significant after FDR correction (p=0.0039 < corrected α value =0.0071), suggesting that the closer the current was to the coast, the further larvae were located from the shore the following day and 44.7±11.74km westwards (Fig 6).

*Theoretical vs real cross shelf larval distributions.* Larvae were located closer to the coast and were much more aggregated at transect 2 than expected if they were evenly distributed, passive particles (Fig. 8, Table 4). Based on our ADCP model, greater offshore dispersal would be expected for passive particles. Observed larval abundances across transect 2 showed a significant fit to a Gaussian function ($R^2$=0.9, p<0.0001, Fig. 8) which produced much lower advective and diffusive distances than the theoretical, mortality-averaged model (Table 3). When time periods of 40h to 8 days were considered, only recently metamorphosed, 40h-old D-stage veligers had a spatial distribution similar to the one observed across transect 2 (Fig. 8). The mortality-averaged alongshore theoretical distribution shows that the source of larvae may be located 12±32km eastward (Table 3).

The periodograms revealed the same temporal patterns for both the alongshore and cross shelf flow at the three depths tested (Table 3). Fluctuations from 5 to 6.33 days explained most of the variability of both flows (Fig. 10). The M2 tidal constituent (12.4 hours) was not apparent in any of the periodograms (Table 3)

DISCUSSION

Our results show that, consistent with other studies, spatial distributions of mussel larvae over the Agulhas Bank were primarily restricted to very nearshore waters (less than 10km from the coast). We also suggest that larval behaviour, consisting of active vertical positioning following light and turbulence cues, helps maintaini the inshore position of larvae over relatively short time scales (order of days). These types of vertical migrations are especially important for the population connectivity in a region that is influenced by the inshore intrusion of the Agulhas Current.

*Mechanisms of active nearshore retention*: Previous studies have suggested that the passive retention through the coastal boundary layer can explain nearshore larval distributions (Wolanski, 1994; Siegel et al., 2003; Largier, 2003). Our moored ADCP data indicate that offshore advection is substantial on the southeastern Agulhas Bank and does not match the expected distributions if the larvae were passive. Theoretical passive distributions would only match the real spatial pattern if unrealistically high mortality rates were used to correct the advective and diffusive lengths. These high mortality rates would change the age structure of the larval population across transect 2 such that all larvae would be recently metamorphosed D-stage veligers (Fig. 8)

Our comparisons represent a rough approach, as both real and predicted larval distributions have been calculated from data for a single location (Eastern Algoa Bay). Spatial variability in the flow along the coast could affect our estimates. According to the mortality-averaged alongshore theoretical distribution, most of the larvae found at transect 2 come from 12±32 km eastwards (Table 4) and upstream conditions are not known. These distances resemble the alongshore spatial correlation scales of 20-40km estimated for coastal currents (Davis 1985; Dzwonkowski 2009). This match suggests that, during their pelagic life, larvae found at transect 2 may have experienced very similar currents to the ones we used for our calculations. In addition, abundance patterns across transect 2 conform with the general decreasing offshore trend observed for the entire Agulhas Bank, , with the notable exceptions of transects 5 and 6. Moreover, maximum abundances were often located even closer inshore than the closest onshore station sampled during this study (unpublished. data), thus underpinning the differences between real and expected abundance patterns. This is consistent with other studies which found high larval abundances at different distances from the shore but always within the coastal boundary layer (Nickols et al., 2013; Weidberg et al., 2013). On the other hand, the already large spatial scales of diffusion and advection obtained from the theoretical model may be underestimated, as the magnitude and variability of the flow would increase tens of kilometers offshore from the position of the current meter (Largier, 2003). In addition, our diffusive distance estimates may be conservative as they were calculated using a decorrelation time scale of only 0.5 days and this parameter rapidly increases offshore (Siegel et al., 2003). Thus, the striking difference found between real cross shelf distributions and theoretical predictions is likely to have been underestimated.

The remarkable difference between calculated and observed distributions most likely reflects the influence of active behaviour of larvae on their own spatial patterns. Different kinds of swimming patterns triggered by environmental cues may drive active locomotion of coastal zooplankton, mainly in the vertical axis (Genin et al., 2005). For mussel veligers, active movement must primarily occur vertically through the water column as they are too slow to contend with horizontal flow: swimming speeds typically reach only a few mm.s$^{-1}$ (Mileikovsky, 1973; Chia et al., 1981). Indeed, our results suggest that mussel larvae perform vertical migrations in response to turbulence and light: larvae were located deeper as both parameters increased.

Turbulence may passively increase sinking velocities of phytoplankton cells through the water column (Ruiz et al., 2004). For marine invertebrate larvae, however, active avoidance of turbulence has been proposed as an effective mechanism to escape from the surface layers of the ocean as when the wind rises. In this way, larvae can avoid wind driven transport (Pringle, 2007). The threshold level of turbulence that triggers this behavioural response is in the order of $10^{-7}$ W kg$^{-1}$ for copepods (see Pringle 2007 for review) and sea urchin plutei larvae (Roy et al., 2012), which may be slightly lower than levels recorded during our sampling. Veligers are also known to retract their velum and sink under turbulent conditions (Young, 1995; Fuchs et al., 2004; Fuchs et al. 2007). Moreover, For *Mytilus edulis,* Fuchs and Di Bacco (2011) found that avoidance occurred at $8.3 \times 10^{-6}$ W kg$^{-1}$ for larval lengths between 171 and 256 µm. We collected smaller larvae (70-200µm, Porri et al., 2014) than those tested by Fuchs and Di Bacco (2011), implying that the larvae from our study should be affected by weaker turbulence consistent with the levels recorded during the cruise.

Many zooplankters undergo diel vertical migrations (Raby et al., 1994), which is thought to reduce offshore transport by decreasing the time larvae spend in the surface Ekman layer (Giraldo et al., 2002; Marta-Almeida et al., 2006; Carr et al., 2007). By correlating MDD and SPAR, we find that the depth of *Mytilus edulis* in the water column is affected by the amount of incoming light (Table 2). This result is not in agreement with studies of copepods in the field (Lagadeuc et al., 1997) or *Mytilus edulis* larvae in the laboratory (Fuchs and Di Bacco, 2011), which suggests that light only has a significant effect in the absence of turbulence. Some decapod larvae exhibited consistent larval vertical migration in the field but not in the laboratory, where light cycles were simulated, suggesting that turbulence and/or food availability rather than light may be the trigger for this behaviour (Miller and Morgan, 2013). Interspecific depth regulation mechanisms are likely to occur, especially for faster and much more active decapod larvae (Queiroga and Blanton, 2004; Miller and Morgan, 2013). Even considering their limited swimming abilities and supposing mean sustained speeds of 0.3 cm/s (Mileikovsky, 1973), it would take less than 7 hours for mussel larvae to go through a water column of 71.5m, the mean depth of the coastal waters where they were actually found (71.5±36.76m, n=36 stations). This time interval may allow diel vertical migration and is much shorter than the inertial period at mid-latitudes (around 17h), thus veligers may be able to "escape" the turbulent surface layer before the onset of Ekman transport (Pringle, 2007).

Despite the indirect evidence of active larval behaviour, we cannot rule out the influence of purely physical processes on larval distributions. For the coastal stations, mean larval depths were shallower with increasing upwelling indices (Table 2, Fig. 5), indicating that the upward flow displaced larvae passively towards the surface during active upwelling. This effect became significant on time scales of 5 days (Table 2, Fig 5). This temporal lag is consistent with the evolution of flow patterns under the forcing of wind-driven upwelling; a frictional period of at least 4 days is required to generate the surface and bottom Ekman layers and the vertical currents between both, typical of a fully developed upwelling episode (Dever, 1997; Austin and Lentz, 2002). Thus, the effects of larval behaviour on spatial distributions are time constrained. There may be a primary active response of larvae avoiding wind driven turbulence and light, but on time scales much longer than the inertial period, vertical Ekman circulation develops and veliger transport may become passive. The time scales at which wind driven turbulence and Ekman transport appear to have acted on larval spatial assemblages (2 and 5 days, respectively) are consistent with those at which both processes varied with time (1 and 5 days, respectively, Table 3).

Our results suggest that tidal amplitude failed to explain larval distributions through the water column because it did not affect circulation patterns. A tidal signature was absent from the flow time-series at Bird Island (Table 3). In fact, fluctuations in currents occurred over periods which matched the periodicity of the wind speed (around 5 days, Table 3), indicating that coastal hydrography was mainly influenced by meteorological conditions. Similar periodicities were obtained in earlier studies for the wind systems passing through Algoa Bay (Schumann et al., 1991) and over the Agulhas Bank (Largier et al., 1992). It is more likely that tidal forcing may determine water column dynamic structure and in turn larval distributions in shallow, enclosed systems like estuaries (Epifanio & Garvine 2001).

By allowing nearshore retention, active vertical migrations triggered by turbulence and light may also alter connectivity patterns along the coast. When included in circulation models, larval behavior in the vertical axis significantly changed dispersal kernels for the whole meta-population (Robins et al., 2013). Similar models have been tested successfully elsewhere in the world, highlighting the importance of active vertical migration by larvae to recruitment dynamics (Domingues et al., 2013) and larval aggregation (Yanicelli et al., 2012) in crustaceans.

*Agulhas-mediated offshore transport:* The median distance offshore for the larval population was found at tens of kilometers from the coast at transects 5 and 6, in marked contrast to the patterns observed at the rest of the cross shelf transects. This offshore plume in larval concentrations matched in space (St Francis Bay) and time (4-5/09/2010) the occurrence of an Agulhas Current solitary meander that propagated downstream over the continental shelf (Jackson et al., 2012; Porri et al., 2014). This suggests that under such powerful physical forcing, larval behaviour failed to achieve onshore retention because, mussel veligers inside the meander (transects 5 and 6) continued to migrate in the vertical axis following light and turbulence cues. Mean larval depths at transects 5 and 6 conformed to the linear trends with SPAR and ε-2days that were valid for the rest of the dataset. Thus, active behaviour did not fail but was simply not effective in terms of onshore retention under the flow structure imposed by the Agulhas Current. Cross shelf flow profiles show that currents were quite uniform in the vertical axis along transects 4, 5 and 6, a feature already observed in the general structure of the Agulhas Current (Bryden et al., 2005). Vertical migrations may lead to nearshore retention if larvae move to layers where offshore transport does not occur. By eliminating those layers and imposing uniform, fast flow throughout the water column, the Agulhas Current overrides the effects of vertical larval migration on coastal retention. The overall result may be the entrainment of larvae in high numbers into the current (densities higher than 10 000 ind. $m^{-3}$, Fig 3) and their offshore export. As meanders entering the inner shelf are unlikely to occur westwards from St Francis Bay (Lutjeharms, 2006), larvae advected seawards are unlikely to be returned to the shelf and are lost to the coastal system. Hence, these events may constitute real losses for the adult population in terms of recruitment and will affect replenishment and connectivity patterns. (Porri et al., 2014). Agulhas- mediated export events are likely to represent larval losses for these populations and probably influence mesoscale connectivity patterns. Such events are relatively frequent (4-6 times per year, Roualt and Penven, 2011) and, among other factors, they could contribute to the fact that mussel recruitment rates along this coast are markedly lower (sometimes by orders of magnitude) than for northern temperate areas (McQuaid and Phillips, 2006) and indeed, the West coast of South Africa (Reaugh-Flower et al., 2011).

Our regression analyses show that the offshore advection of mussel larvae by the Agulhas Current may occur on a fixed spatio-temporal scale given the physical characteristics of the current (Fig 6). Distance from the current was correlated with MLDO only when a lag of one transect (1 day, 44.7±11.74km) was introduced into the model. This becomes evident in the cross–shelf profiles; the closest position from the inshore edge of the Agulhas Current to the shore was recorded at transect 5 (50km), while the maximum MLDO was found at transect 6. Thus, the effect of encroachment by the current on the cross shelf distributions of larvae becomes clear the next day and about 44.7±11.74km downstream. Theoretical spatio-temporal scales for this process seem to match real observations. The distance between the nearshore end of transect 5 and MLDO at transect 6 was approximately 70 km, which is consistent with the net distance theoretically travelled by larvae trapped in the edge of the Agulhas Current at typical speeds of 1m s$^{-1}$ during 24 hours (86 km).

Powerful, but relatively infrequent mesoscale hydrographic features which resemble Agulhas Current Meanders have been shown to affect larval distribution and settlement in several coastal systems. Entrapment and transport by oceanic eddies of reef fish larvae, much faster swimmers than mussel veligers, was observed off Hawaii and Florida (Lobel and Robinson, 1986; Graber and Limouzy-Paris, 1997). Prolonged and intense warming events, associated with the eddy field in the Humboldt Current, significantly changed the spatial scales of recruitment along the Chilean coast (Narvaez et al. 2006). Similarly, although variability in normal upwelling dynamics may not affect settlement rates on the west coast of North America (Fisher et al. 2014), large relaxations characterized by poleward boundary currents and the entrainment of low salinity waters resulted in offshore larval transport along central California (Morgan et al. 2012). Contrary to our results, which point to active vertical positioning of mussel veligers within the Agulhas Current Meander, larvae may suppress vertical migration during these upwelling relaxations events, thus avoiding the surface low- salinity lens (Morgan et al 2012).

Mussel larvae were also found tens of kilometers from the shore at transects 9, 10 and 12, where they were at low densities and showed less spatial structure. These patterns may not be the result of offshore exportation process but rather of aggregations at the crest of shoaling internal waves affecting the seaward "tail" of the Gaussian cross shelf larval distributions. The presence of these physical structures across transect 12 was clear from temperature and flow profiles (Jackson et al., 2012). The relatively high larval concentrations close to the surface could depend on the presence of internal waves which can accumulate plankton at the surface due to their characteristic convergent flow (Lennert-Cody and Franks, 1999; Pineda 1999).

**Conclusions:** Our findings show that effective connectivity among populations on this coast is mediated by larval mobility and by the ineffectiveness of this behavior when the Agulhas Current intrudes into the nearshore environment. Frequent nearshore hydrographic features can be exploited by larval behaviour to produce effective connectivity in the absence of the Agulhas Current. Under forcing by the Agulhas, however, it is likely that connectivity between large portions of the larval pool and adult habitats is entirely disrupted and larval behaviour no longer matters. Consequently, for effective connectivity, larvae need an alignment of hydrographic processes as well as behavioural ability, both of which can be disrupted by meso-scale (i.e. Agulhas current) features. Our results highlight the need for consideration of complex, region-specific physical dynamics and their interactions with the behaviour of the organisms in order to explain patterns of connectivity in benthic populations.

Acknowledgements

This work is based upon research supported by the South African Research Chairs Initiative of the Department of Science and Technology and the National Research Foundation and it was funded by the African Coelacanth Ecosystem Programme (ACEP II). Funding for JJ was provided by the ACEP II program and the National Science and Engineering Research Council of Canada. The authors thank the captain and crew of the RV Algoa, Ryan Palmer, Dr Jacky Hill, Cornelia Nieuwenhuys, Max Seigal, and Luzuko Dali for assistance and data collection. We would also like to thank Raymond Roman for his help with processing the ADCP data.

# Tables

Table 1. Acronyms for all predictors and dependent variables

| Acronym | Variable | Units |
|---|---|---|
| MDD | Mean depth distribution | m |
| MLDO | Median larval distance offshore | m |
| SPAR | Surface photosinthetically active radiation | W m$^{-2}$ |
| TA | Tidal amplitude | m |
| UPW | Upwelling index | m$^3$ km$^{-1}$ s$^{-1}$ |
| ε | Wind-induced turbulence | W kg$^{-1}$ |

Table 2. Regression analyses with MDD as dependent variable. Models were ordered following the AIC index. Regressions were performed for all the stations where larvae were found (abundances > 10 ind. m$^{-2}$, named All, n=36) and for a subset of coastal stations (first three stations off the coast, named Inshore, n=26). See Results for the meaning of the acronyms. (*<0.05; **<0.01; ***<0.001).

| Model | | AIC | R$^2$ | p value | n° variables | Predictors | Std coefficients |
|---|---|---|---|---|---|---|---|
| All | 1 | 302.367 | 0.374 | *** | 2 | SPAR | -0.502 |
| | | | | | | ε(2days) | -0.314 |
| | 2 | 304.899 | 0.276 | ** | 1 | SPAR | -0.525 |
| | 3 | 311.575 | 0.124 | * | 1 | ε(2days) | -0.352 |
| Inshore | 1 | 208.516 | 0.192 | * | 1 | UPW(5day) | 0.438 |
| | 2 | 209.441 | 0.252 | * | 2 | SPAR | -0.377 |
| | | | | | | ε(2days) | -0.343 |

Table 3. Dominant periods for each variable revealed by Fourier transform analyses and their corresponding periodograms. Winds were measured hourly at Port Elizabeth weather station from 15 August to 15 September 2010. Currents were measured every 20 minutes at the ADCP currentmeter in Bird Island from 1 August to 30 September 2010.

| Variable | Dominant period |
|---|---|
| Wind, cross shore component | 110h |
| Wind, alongshore component | 110h |
| Upwelling index | 110h |
| Wind induced turbulence | 24h |
| Cross shore currents, 6.3m depth | 114h |
| Cross shore currents, 12.8m depth | 114h |
| Cross shore currents, 28.8m depth | 114h |
| Alongshore currents, 6.3m depth | 114h |
| Alongshore currents, 12.8m depth | 152h |
| Alongshore currents, 28.8m depth | 152h |

Table 4. Advective and diffusive distances in km ($L_A$ and $L_D$, respectively) from the theoretical, moored ADCP models and from a Gaussian fit to real larval abundances across transect 2. Standard error for each value is shown.

|  | $L_A$ | $L_D$ |
|---|---|---|
| Mortality-averaged, whole period | 42.46±3.21 | 23.45±4.20 |
| Mortality-averaged, whole period, alongshore | 12.74±4.75 | 32.72±4.86 |
| 40h | 3.77 | 4.26 |
| 3 days | 29.30 | 20.46 |
| 4 days | 55.86 | 26.00 |
| 6 days | 57.71 | 30.30 |
| 8 days | 74.74 | 34.46 |
| Real | 5.75±18.89 | 6.96±12.08 |

**Figure legends**

Figure 1. Grid of stations sampled from 31 August to 13 September 2010. Black dots denote stations where both CTD profiles and plankton-pump data were collected while grey dots show stations where only CTD profiles were obtained. The 100m isobath is represented in dark grey. Red inverted triangles show the position of the Agulhas Current according to a temperature threshold at each transect during sampling. White dots represent meteorological stations were wind data were collected. Tidal amplitude was also measured at those represented with a dark cross. The asterisk shows the location of the moored ADCP current meter at Bird Island. The acronyms stand for the names of the weather stations and other places mentioned in the text. From West to East: Cape Agulhas (CA), Struisbaai (SB), Mossel Bay (MB), Plettenberg Bay (PB), Cape St Francis (CS), St Francis Bay (SB), Cape Recife (CR), Port Elizabeth (PE), Algoa Bay (AB), Cape Padrone (CP).

Figure 2.Contour cross shore profiles of temperature/salinity, cross shelf currents and larval abundances for transects 1, 2 , 3 and 4.. Current profiles were only obtained from depths below 35m. In the case of the currents, positive (red) and negative values (blue) point to landwards and seawards flow, respectively. For temperature and salinity, isotherms are represented every 1ºC while color contours show variations in salinity. Larval abundances lower than 10 individuals m$^{-3}$ are represented with open circles. White surfaces indicate the absence of data points. Inverted triangles show the position of the stations at which each variable was measured.

Figure 3. Contour cross shore profiles of temperature/salinity, cross shore currents and larval abundances for transects 5, 6 , 7 and 8. See Figure 2 for details.

Figure 4. Contour cross shore profiles of temperature/salinity, cross shore currents and larval abundances for transects 9, 10 , 11 and 12. See Figure 2 for details.

Figure 5. Simple linear fits to mean depth distributions using different independent predictors (n=36). Grey dots represent the stations located inside the Agulhas Current meander at transects 5 and 6 where offshore exportation of larvae was observed. For the upwelling index averaged over 5 days, only the three most coastal stations for each transect were included (n=26). Although non-significant, the linear fit for tidal amplitude is also represented. See Table 2 for the significance and parameters of the fits.

Figure 6. Significant simple linear fit to MLDO using distance to the Agulhas Current with a lag of 1 transect as independent predictor. See Results for acronyms and significance of the fit.

Figure 7. Sea surface temperature satellite imagery (http://www.remss.com/sst) for transects 1, 2, 3 and 8 (A: August 31, B: September 1, C: September 2, D: September 7). Dark cross shore lines show the orientation of the transects with a perpendicular bar at their offshore end representing the location at which the inner border of the Agulhas Current was presumably placed (red contour line based on temperature thresholds, see

Material and Methods). Black dots point to the stations where larvae were sampled. PE and PB stand for Port Elizabeth and Plettenberg Bay, respectively.

Figure 8. Theoretical normal distributions based on the flow measured at the ADCP current meter in Bird Island under a passive particle-assumption: averaged for the whole larval period and corrected by mortality (grey line); and calculated for specific periods of 40h, 3 days, 4 days, 6 days and 8 days (black, red, green, orange and blue dotted lines, respectively) . The Gaussian fit to real larval abundances at transect 2 is represented with a dark line.. Black dots represent mean larval abundances at each station across transect 2 with bars showing their respective standard deviation. This fit was performed pooling data from the three different depths (n=21). See Table 4 for significance and parameters of the functions.

Figure 1.

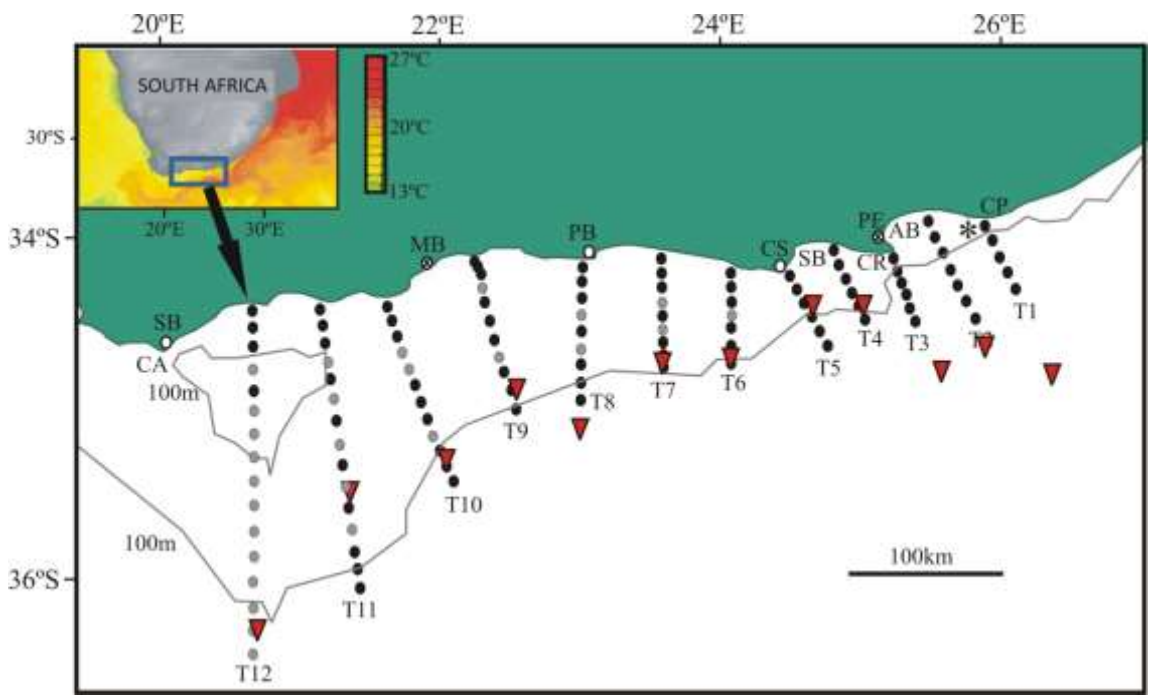

Figure 2.

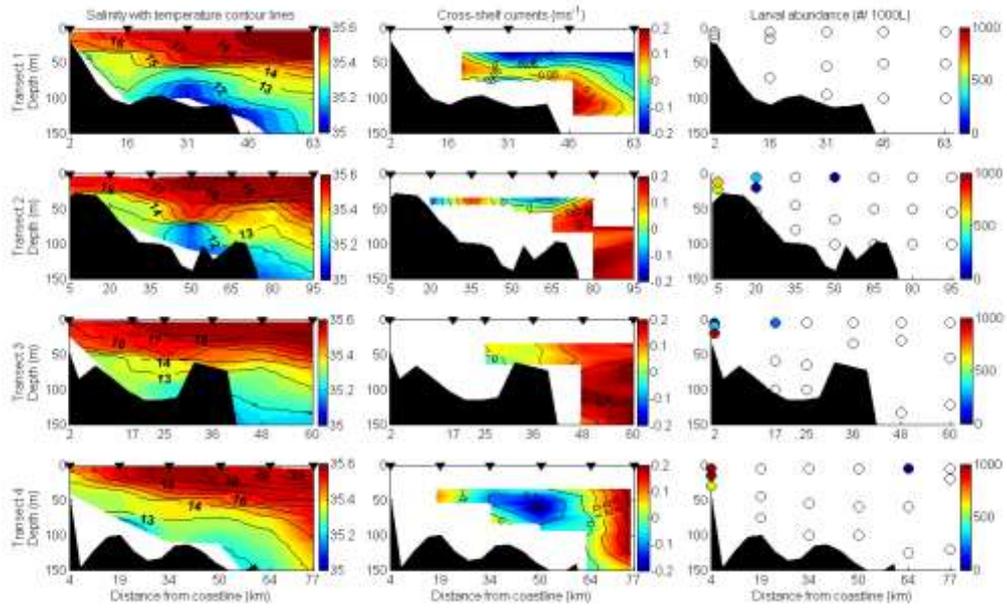

Figure 3

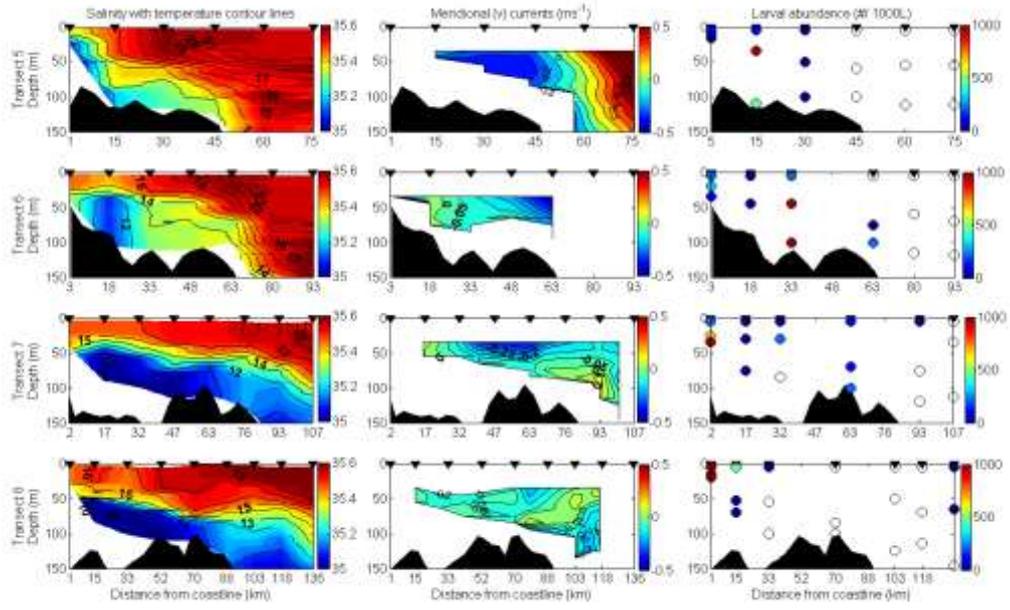

Figure 4.

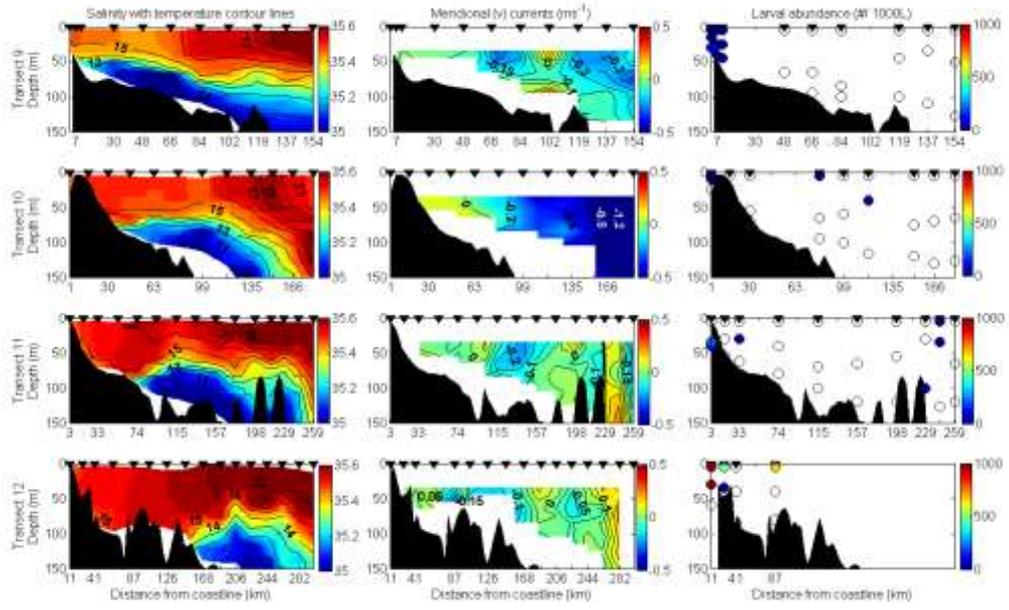

Figure 5.

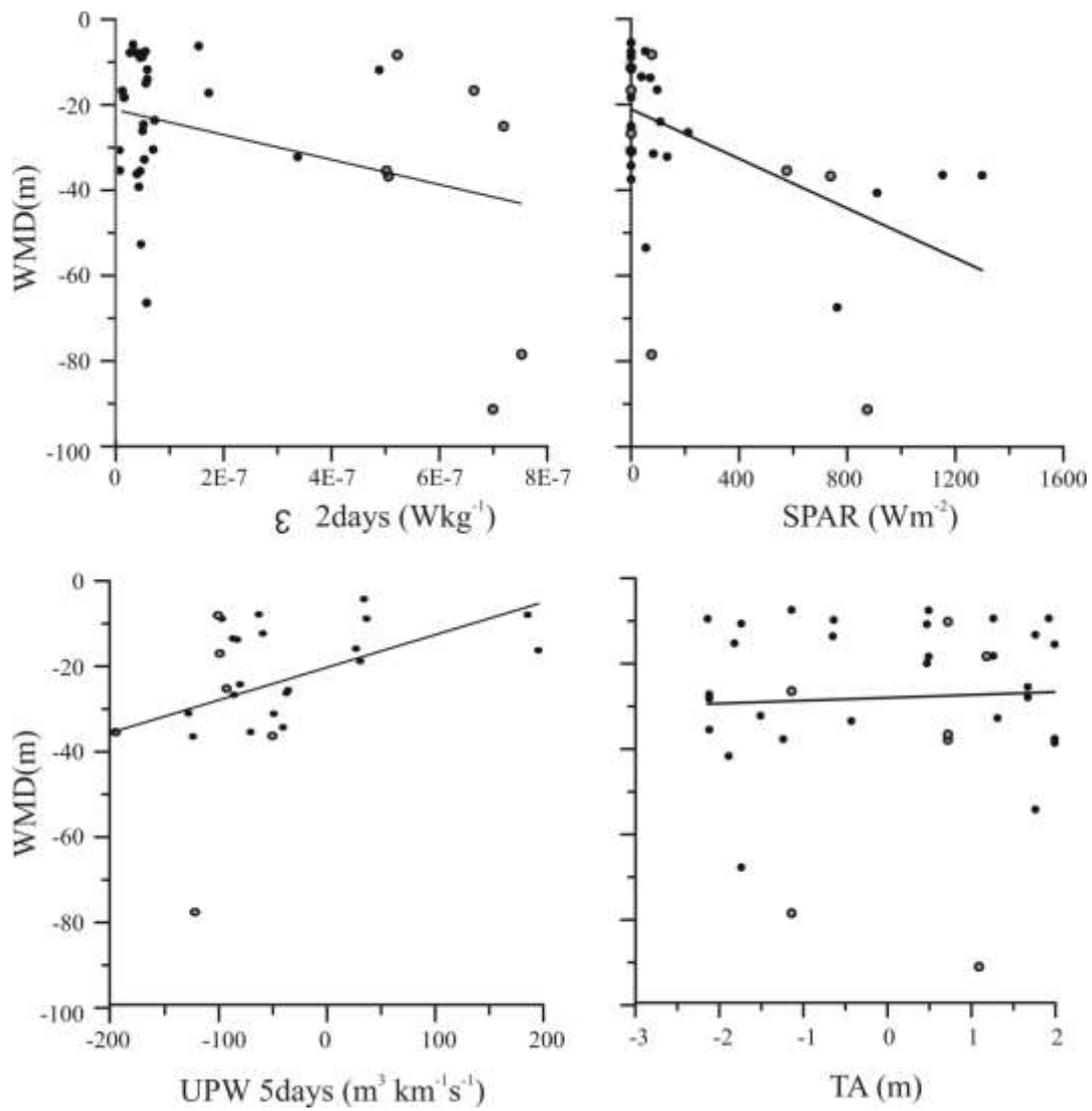

Fig. 6

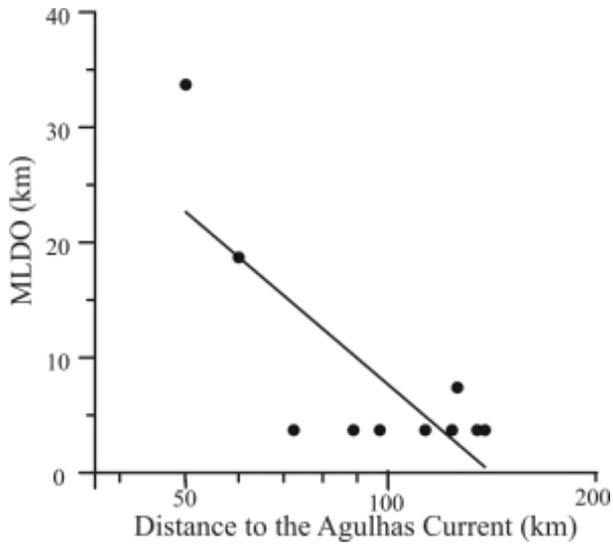

Figure 7

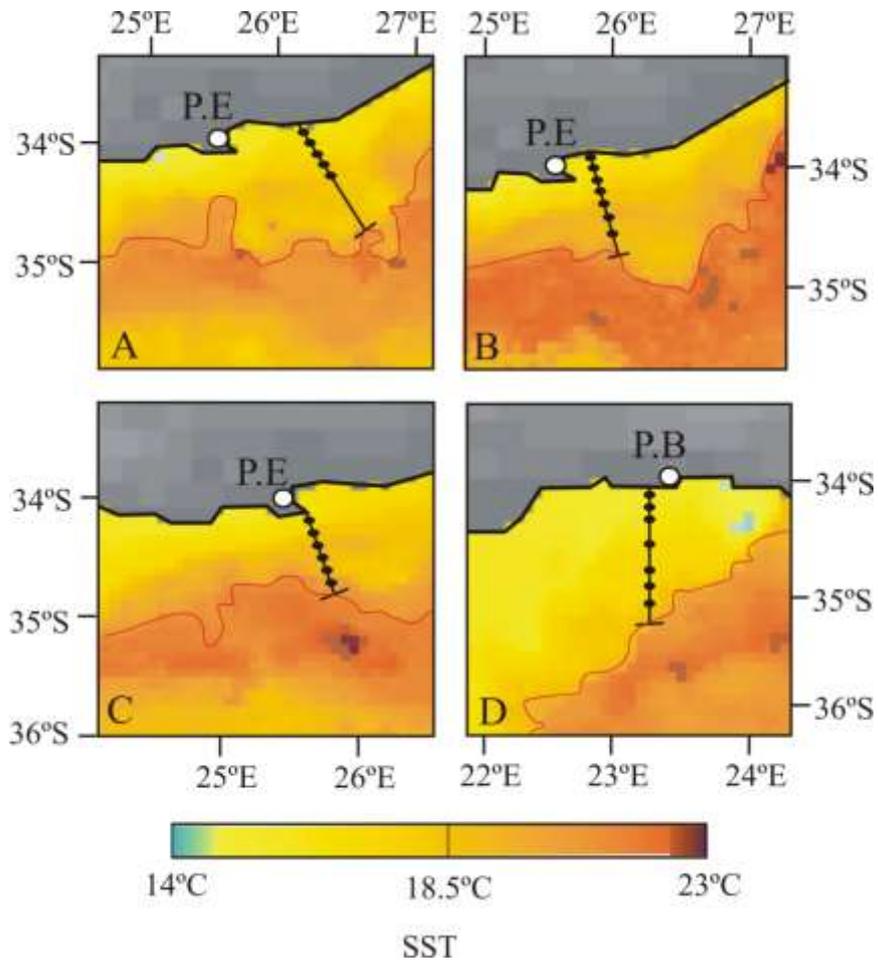

Fig. 8

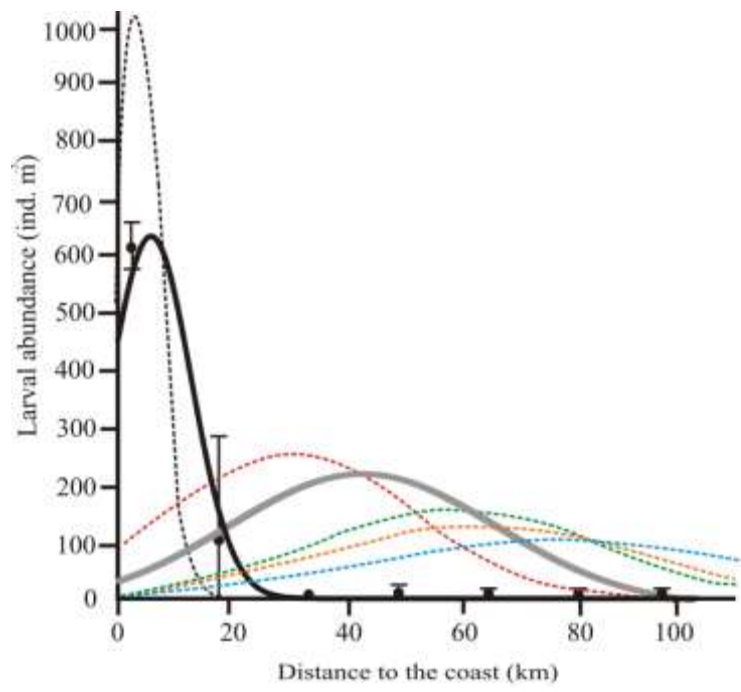